\documentclass[onecolumn,10pt]{article}
\usepackage{times}
\begin{document}
\global\def\refname{{\normalsize \it References:}}
\baselineskip 12.5pt
%
%
%
\title{\LARGE \bf Mathematical Tool of Discrete Dynamic Modeling of
Complex Systems in Control Loop}

\date{}

\author{\hspace*{-10pt}
\begin{minipage}[t]{2.7in} \normalsize \baselineskip 12.5pt
\centerline{\large A\small RMEN \large B\small AGDASARYAN}
\centerline{V.A. Trapeznikov Institute for Control Sciences, Russian Academy of Sciences}
\centerline{Profsoyuznaya 65, 117997 Moscow}
\centerline{Russia}
\centerline{abagdasari@hotmail.com}
\end{minipage} \kern 0in
%
%
\\ \\ \hspace*{-10pt}
\begin{minipage}[b]{5.6in} \normalsize
\small In this paper we present a method of discrete modeling and analysis of multi-level dynamics of complex large-scale  hierarchical dynamic systems subject to external dynamic control mechanism. In a model each state describes parallel dynamics and simultaneous trends of changes in system parameters. The essence of the approach is in analysis of system state dynamics while it is in the control loop. 
\\ [4mm] {\it Key--Words:}
Discrete Modeling, Dynamics, Control Loop, Complex Systems, Large-Scale Hierarchical Objects  
\end{minipage}
\vspace{-10pt}}

\maketitle

\thispagestyle{empty} \pagestyle{empty}
%
%
\section{Introduction}
\label{S1} \vspace{-4pt}

When modeling and analyzing of control and dynamic processes in complex multi-component large-scale systems it is necessary to operate with multiple state coordinates. This is caused by (1) the fact that complex large-scale system behavior is influenced by a number of factors of various nature which leads to large amount of system parameters, indicators, and variables; (2) lack of sufficient quantitative information (incompleteness, uncertainty) on the processes that influence system development, especially, for systems and objects belonging to weakly-formalisable ones. Therefore, a peculiar approach, which will allow to take into account all essential diverse factors that determine system activity and behavior under the influence of external control actions, is needed. The modeling technique developed allows one to cope with the above mentioned issues. The basic features of the proposed approach are the following:
\begin{enumerate}
	\item As a canonical model of control object we consider multi-level dynamical systems consisting of a set of autonomous elements (subsystems) with own individual (local) and general corporate (global) problems and goals;
	\item The external dynamic control mechanism in a system is considered as a set of control actions initiating multi-level state dynamics of control object;
	\item The long-term databases and monitoring that characterize the changing of parameters and indicators can be used as main source of information about system development and control problems. Databases and other statistical material facilitate observing for the changes of parameters at different time intervals, which has an extreme promise for understanding the global regularities in system dynamics. Monitoring includes observation of the current situation around the system. The processes under monitoring are interpreted in the form of state dynamics and estimation, and tendencies of system development as well. The system goals are formulated as consistent dynamics of these processes. Monitoring of the present situation enables (a) discovering new factors and parameter estimations influencing the system development, (b) establishing possible new or desirable states and goals. In this case the model is updated;
	\item Due to the hierarchical structure of parameter set, a multi-level (hierarchical) control loop based on a set of independent closed control loops of lower levels is constructed. The basic criteria for multi-level control loop efficiency are consistency and time-event coordination of attainable states (goals) and of dynamical properties of system parameters. 
\end{enumerate}	
In our approach a large-scale hierarchical system is understood as a combination of distributed in time and space interacting subsystems that organize separate hierarchical levels. On each level a subsystem is assumed to be described in corresponding space of parameters and variables, some of them are so called polymorphic that equally applicable for objects at different levels of hierarchy. On each hierarchical level the system has its local goals.
	
A control problem of system development is considered as a construction of controlling scenario realizing a time-event coordination of control actions to achieve control goals of subsystems at different hierarchical levels and at the same time to implement global system goals.

In a whole, system functioning efficiency depend not only on the "top-down" influence but on the "bottom-up" response as well, i.e. on the consistent behavior of all system elements.

So, the aim of this paper is to propose the method for constructing discrete models of complex hierarchical dynamic systems subject to external hierarchical dynamic control mechanism and their problem-oriented interpretation. The method includes:
\begin{enumerate}
	\item Creating multivariate multi-level hierarchical structural model based on system analysis;
	\item General methematical formalisation;
	\item Constructing hierarchical dynamic graph model to solve the system development control problems and to analyze system dynamic characteristics related to the attainability of desirable states and goals;
	\item Specializing the model to the scenario control schemes of complex systems.
\end{enumerate}

\subsection{Problem Formulation, General Features and Description of Model}
\vspace{-4pt}

In connection with what has been said above the key requirements to the model of system development control are the following:
\begin{enumerate}
	\item Representation of interrelations between local and global goals;
	\item Consistency and efficiency of control actions in the process of goals achievement (at the level of required values and dynamics of parameters);
	\item Time and event ordering of needed control actions;
	\item Problem-oriented significance of each control action.
\end{enumerate}
The basic complexity of the problem under consideration is:
\begin{enumerate}
	\item to make consistent and to coordinate a set of problems and conflicting goals in long-term system development
	\item to analyze tendencies, shifts, and proportions in parameter changes, and classification of control objects;
	\item to make consistent practically unlimited number of dynamical processes.
\end{enumerate}
This raises the problem of choosing the principles which to a maximal extent reduce all the variety of dynamical interrelations when modeling a large-scale hierarchical system to clear and logic constructions. The following basic principles of system modeling have been chosen:
\begin{enumerate}
	\item Systematic and necessary use of hierarchy. In our case, the hierarchy is used for representation of control domain and its qualitative characteristics (polymorphic parameters). The principle of hierarchy allows us to distinguish the essential interrelations for aggregation and scaling (recount) of dependent parameters, and also helps structuring the problem domain of an object;
	\item Use of the concept of state as system indicators. The notion of state is used as a mean for description of combinability of values of various parameters, logically coupled and uncoupled;
	\item Use of state diagrams in control loop. The efficiency of control actions and comparison for efficiency of different sets of control actions are formulated on the basis of state diagrams.
\end{enumerate}
To cope with the problem of dimension due to the growth of the model, the states are considered not as a combination of parameter values or its ranges but as a combination of parallel trajectories of parameter changes (principle of getting the integrated notions of pseudo-organisational system). The description of trajectories includes not only primitive growth/drop of parameter values but also some typical paths of changes of the parameter value ranges with stable and unstable cycles. We include in one state a set of statical state characteristics on time interval and get an aggregated characteristics for practically unlimited number of parameters simultaneously. In combination with the principle of hierarchy this is a convenient tool for description of interrelations between tendencies, shifts and proporions in changes of parameter values related to the objects at the different level of system. This, in turn, serves as a basis of analysis for consistency of dynamics of various hierarchical parameters of system elements. The principles of modeling given above enable us to formulate a universal model of complex large-scale hierarchical dynamic system in control loop.

\section{State Diagrams as Formalism for Representation of System Development Dynamics}
\label{S2} \vspace{-4pt}

In this section we describe a model of development of complex hierarchical objects. We introduce hierarchical state diagrams as a tool for formal representation of system parameter dynamics. The dynamics of parameter changes of objects is the main body of the model.

\subsection{Generalized Scheme}
\vspace{-4pt}
The system approach to the modeling of complex systems supposes the analysis of interconnected processes for as many components as it possible. To satisfy this requirement we organize the modeling so that the objects development is manifected in the form of state dynamics which characterizes both the system as a whole and its components. The formalisation of state processes and of dynamics of general and special parameters in the form of mathematical mappings underlies the proposed method. In a generalized form the model of system development and hierarchical control is represented as follows.

Let $F_{ij}$ be a set of control actions for $ij$-th subsystem, $F_{ijk}$ a subset corresponding to $k$-th state, $[0,T^*]$ the control time interval, and $f(i,j,t)\subseteq F_{ij}$ the control action on $ij$-th subsystem at the moment $t\in[0,T^*]$. Then the control process is described by the vector-function $$f(t)=(f_1(i,j,t),f_2(i,j,t),...,f_n(i,j,t))$$ in control space $\prod^{n}_{1}{F_i}$ of Cartesian product of sets of control actions on different subsystems. We consider that $f(i,j,t)$ influence uniquely on subsystem state and on the value of its efficiency criterion. Let $s(i,j,t)$ be a process of state changes, and $w(i,j,t)$ a process of efficiency criterion changes for $ij$-th subsystem on control time interval $t\in[0,T^*]$. The the vector-functions $$s(t)=(s_1(i,j,t), s_2(i,j,t),...,s_n(i,j,t))$$ and $$w(t)=(w_1(i,j,t), w_2(i,j,t),...,w_n(i,j,t))$$ describe the attainable configurations that represent the efficiency of control process $f(t)$ at the moment $t\in[0,T^*]$.

\subsection{Principles of Construction of Development Model of Complex Hierarchical System in Control Loop}
\vspace{-4pt}
The scalability, i.e. the simultaneous representation of goals and development character of various object component, is very important when constructing a model of large-scale system development. Two factors play an important role in providing the scalability. The first one is the systematic use of hierarchical principle for representation of control object, control system, and system parameters. The second one is to establish polymophic parameters, equally applicable for object at different levels of hierarchy. Polymorphic parameters in hierarchical models enable one to turn from control at the object level to control at the level of object classes, and also from individual models to integral models of arbitrary level of generalization. In this case, the problem of modeling of complex large-scale hierarchical system development can be reduced to the analysis and interpretation of long-term dynamics of polymorphic hierarchical parameters of hierarchical object. The most important properties of the model are:
\begin{enumerate}
	\item object development models at each level are sufficiently autonomous. This provides a sufficient degree of decomposability and therefore flexibility of large-scale models construction;
	\item modeling objects are not only separate components but also classes of components having common development goals;
	\item models of system components are turned out to be information compatible, outputs of one component can serve as inputs for another component.
\end{enumerate}
Except for scalability, another important requirement to development models of complex hierarchical objects is to use abstract notions for qualitative description of long-term dynamics of parameters. The basic idea concerning the abstract representation of process dynamics is to use state diagrams.The continuous time interval $[0,T^*]$ is divided into parts. On each part a process is described by a state, and interaction between parts is described by state diagram. A state of hierarchical object is defined as a situation which is characterized by a set of states of object components. Each state is a set of characteristic trajectories of parameters changes. Deformation of trajectory character on transition from one part to another formalises the state change. The state change can be used for estimation of direction, efficiency and quality of control actions. In this way we get a qualitative image of state dynamics which is essential for control goals. This helps representing the real and desirable characteristics of control object, their properties, structural and functional interrelations. The approach to the representation of system states has the following properties:
\begin{enumerate}
	\item aggregates the parameters and, therefore, simplify the system modeling;
	\item formalises the information gathering and estimation for getting the integrated and local evaluation of hierarchical system;
	\item forms the basis for analysis of system dynamics and facilitates the study of a number of aspects of dynamical process in a unified way.
\end{enumerate}
We need to construct a formalised model which is based on universal scheme of decomposition of hierarchical object development model. Controllable objects belonging to a hierarchical set $W$ are estimated by polymorphic parameter of hierarchical structure $I$. Each component is represented by state diagram from a set $D$ of state changes (real, desirable, predictable). The diagrams from $D$ that correspond to the objects of one level of hierarchy represent sequential processes of state changes. The diagrams of objects of higher level of hierarchy represent the corresponding (parallel, taking place at the same time moment) states of higher level.

\subsection{Formalised Scheme of Construction of Development Model of Complex Hierarchical System with Polymorphic Parameters}
\vspace{-4pt}
In this subsection we give a scheme for construction of development model of objects of one level of hierarchy. The scheme is a basic algorithmic step of construction of development model of complex hierarchical system with polymorphic parameters.

Let us denote the set of objects of one level of hierarchy $W'$. The scheme is divided into four stages.

The \textit{first stage} includes the preliminary study and consists in establishing the parameters with parallel dynamics, which characterize an arbitrary object from $W'$. At first stage we choose the set of parameters and form the graph representation of their parallel dynamics at given time interval. Using the graph representation we compare the character of parameters changes of object under study.
The \textit{second stage}, the stage of dynamic parameters estimation, consists in getting the comparative dynamical characteristics of polymorphic parameters for different objects from $W'$, and in extrapolating the dynamics of parameter values for arbitrary object with simple relationships, which describe the essence of processes under study. The analysis of parameter dynamics gives answers to the following questions: whether a parameter is a function of time of any standard type, monotone increasing or decreasing, with one or several critical points, whether the function is bounded, whether it has a point of inflexion, or it can be described by a cyclic process. The basic idea of algorithm for recognition of type of the dynamic process consists in estimation of parameter dynamics state. This includes heuristic analysis of a sequence of parameter values $X(1)$, $X(2)$,... and producing the current state process estimate $S(t)=F(S(t-1),X(t-1),X(t))$ in arbitrary time moment. The algorithm is universal and applicable for any parameter, for which values the notion of comparison is defined. A qualitative estimation of the current process of a parameter dynamics enables one to create diverse classification rules for objects from $W'$. This activity forms the \textit{third stage}. Classification rules are given by means of matrices with logical elements. An element $(I,J)$ of matrix, where $I$ is a parameter and $J$ is a class of objects, $J\subseteq W'$, contains a logical formula which determine the current state of process of $I$-th parameter changes. The matrices of this type give rules of one-level classification. However, the most important are rules of hierarchical classification based on the eventual specification of conditions to be satisfied by objects from a class. The notion of state scale and classificator are formal basis for construction of multi-level classification rules.

Let $K=\left\{k_1,k_2,...,k_q\right\}$ be a set of predicates, propositions relating to the parameter values of objects set $\Omega$. The ordered set of predicates $K=\left\{K_1<K_2<...<K_n\right\}$, $T_{K_i}\cap T_{K_j} = \emptyset$, where $T_{K_l}$ is a truth domain of $K_l$, is called a one-level scale (further, scale) if each $K_i$ defines a state $S_i$. It is assumed that predicates and the corresponding states have the same ordering, i.e. if $K_1<K_2<...<K_n$ then $S_1<S_2<...<S_n$. The scale determines the values of parameters and enables us to compare the states of the objects.

We say that a scale $\left\{K_{i_1}<K_{i_2}<...<K_{i_n}\right\}$ is the hierarchical continuation of the scale \\$\left\{K_1<K_2<...<K_i<...<K_n\right\}$ if the predicates $\left\{K_{i_1}<K_{i_2}<...<K_{i_n}\right\}$ are the set of sub-predicates of $K_i$. A hierarchical system of scales is called to be the classificator of objects from $\Omega$ over the hierarchical set of parameters at the time interval $\Delta$. At the \textit{fourth stage} the classificator is used for formal description of dynamic development model of objects $W'$. Formalised scheme of dynamical system description is given in the form of canonical state development model of objects $W'$.

A canonical model of state development of a set of objects is represented at the time interval $[0,T^*]$ by state transition diagram
$$D=\left\{S,K,P,S_0,S^*,\mu_0,\mu^*\right\}$$
$S$ - a set of states ordered by $K$, \\
$S_0,S^*$ - initial and final states respectively, \\
$P$ - a set of arcs; each arc is assigned a time interval $\Delta\in [0,T^*]$ of state transition, if $(S_1,S_2)\in P$ then $S_1<S_2$, \\
$\mu_1,\mu_2,...,\mu_n$ - a sequence of objects distribution over the vertices-states of diagram at $t_1,t_2,...,t_n$ respectively; $\mu_0$ - an initial distribution, $\mu^*$ - a final distribution.

The canonical model formalises the qualitative properties of dynamical system and represents a hypothetical model of development. The description of real development process at arbitrary time interval $[t_i,t_j]$ is based on the use of states of canonical model as objects classificator. To the set of arcs $P$ in canonical model a set $P^0$ is added. $P$ and $P^0$ are called the arcs of state development and the arcs of critical backstep of state, respectively; if $(S_1,S_2,\Delta) \in P$ then $S_1<S_2$, otherwise, if $(S_1,S_2,\Delta) \in P^0$ then $S_2<S_1$. $(S_1,S_2,\Delta) \in P$ means that an object from $W'$, being in the state $S_1$ at the time $t$, transits to the state $S_2$ at the time interval $t+\Delta$. Each arc $(S_1,S_2)\in P\cup P^0$ is assigned the objects counter $\eta$, which changes their state from $S_1$ to $S_2$ at the time interval $[t_i,t_j]$. The counters assigned to the arcs $P$ characterizes the intensity of development processes; the counters assigned to the arcs $P^0$ estimates the intensity of negative processes in object development. Consider the number of objects $N_i$ having a fixed state $S_i$ and the counter $\eta_{ij}$ assigned to $(S_i,S_j)$ as functions of time, $N_i(t)$ and $\eta_{ij}(t)$, on the observation time interval. Introduced parameters enables us to get the information concerning the relation between processes of development and degradation, and the dynamics of processes. This allows us to get a qualitative image of the development processes of dynamical system under study.

The given above four stages comprise the general scheme of study of objects set of arbitrary level of hierarchy as a unified dynamical system.

\section{State Diagrams and Development Models of Complex Hierarchical Systems}
\label{S3} \vspace{-4pt}

This section is devoted to the use of state diagrams for construction of developemnt models of complex objects. The state diagrams technique is a tool for solving a wide range of problems, estimation of control actions, comparison of control actions sets for efficiency, qualitative estimation of processes of system development, and control problem solving.

\subsection{Operations with State Diagrams and their Consistency}
\vspace{-4pt}

In section 2.3 we proposed the analytical description of objects dynamics of one level of hierarchy. Considering the objects from neighbour levels of hierarchy, one can, in principle, create the complexes of development models $\Omega=\left\{\Omega_1,\Omega_2,...\right\}$, $\Omega_i$ are called elementary. The elementary development models enables one to analyze a number of various aspects of hierarchical objects development. However, for large-scale objects the process of analysis of such models may result in a difficult problem. It is preferable to consider the models of state dynamics of multi-component systems, with each subsystem having its own special (local) and general goals. 

The state diagrams technique is universal tool for representation of dynamic developemnt schemes for diverse control problems, not depending on their level and character. This forms a basis for coordination and consistency (concordance) of control problems. In this connection, functional generalization of several elementary models of $\Omega$ and construction of complex development models is of interest. Structural composition of state diagrams of several elementary development models provides a synthesis of complex requirements set to dynamical characteristics of controllable objects. The structural composition holds a central position in the models of hierarchical system dynamics.

Let $D=\left\{D_1,D_2,...,D_i,...,D_n\right\}$ be a set of diagrams to be composed, given at the time intervals $\left\{[0,\tau_1],[0,\tau_2],...,[0,\tau_i],...,[0,\tau_n\right\}$ respectively. Then, we say that for the diagrams $D_i$ the property of consistency holds if the attainability of certain states takes place in a given (prescribed) time-event sequence.

We introduce the basic operations for state digrams in order to give the criteria of their consistency as follows. \\
I. Sequential-parallel composition \\
a) a set of diagrams $D$ forms a linear fragment, sequentially composed, if for their time intervals the following inequality holds $\tau_1<\tau_2<...<\tau_i<...<\tau_n$. \\
b) a set of diagrams $D$ forms a parallel fragment, composed in parallel, if they are defined on the same time interval. \\
II. Generalization \\
To give the criteria for consistency of dynamical systems at neighbour levels of hierarchy we use the Cartesian product of states of diagrams of lower level of hierarchy. In this case, the consistency criteria for state development of dynamical system at neighbour levels of hierarchy is realized by specifying the ordering relation on the subsets of Cartesian product of states of diagrams of lower level of hierarchy. \\
The diagrams composition allows one to formally represent different combinations of complex criteria sets, to perform objects classification, and to solve control problems. Using the consistency rules and operations with diagrams one can model diverse schemes of inter-level relations and influence (effect) of states of lower level diagrams on the processes of higher levels of hierarchy. As a result, a certain value is produced at the output of the highest level. This value is considered as a response, reaction, of the whole hierarchical network (model) on the values of input parameters. 

\subsection{Model of Controllable Development}
\vspace{-4pt}
The various approaches to the control processes require consideration of development models, in which the representation of controllable dynamics of hierarchical object initiated by input signals comes to the forefront. The model of controllable development is vased on the following principles:
\begin{enumerate}
	\item selecting the control actions that influence the controllable system; this is important for autonomous construction of control scenario and for flexible modification of the model to alternative control scenarios
	\item taking into account the states that has been attained on the previous control stages (system state history); this provides a succession of multi-stage control scenarios
	\item comparing with the results of alternative control scenarios; this provides basic arguments upon estimating the efficiency of control scenarios.
\end{enumerate}
The model of controllable development illustrates the key dynamic characteristics depending on whether or not the control actions corresponding to the current states are performed. In this sense, the model of controllable development is constructed in the form of hypothesis "what if...".

A hypothesis is defined by state transition diagram $$D^H=\left\{S,P,S_0,S^*,X\right\}$$
$S$ - a set of states, \\
$S_0$, $S^*$ - an initial and final states respectively, \\
$X$ - a set (alphabet) of input control symbols, \\
$P$ - a set of arcs; $P=P_1\cup P_2$, $P_1\cap P_2=\emptyset$ \\
$P_1$ - the subset of arcs of state transitions initiated by input symbols, \\
$X\Leftrightarrow P_1$ - a correspondence  that determine for each input symbol the state transition initiated by the symbol, \\
$P_2$ - the subset of arcs of state backstep in the absense of input symbols.

To model and analyze the connections between different subsystems we introduce a mechanism of win/loss that other subsystems can obtain depending on the state of each element.

Let us denote the input alphabet (input symbol set) $X=\left\{\cup X\right\}$, the set of arcs $P=\left\{\cup P_1\right\}$, of a set of state transition diagrams. 

We define the mechanism of after-effect by splitting $P$ and $X$ into two subsets $(Z,U)$ and $(X^Z,X^U)$, respectively. The arcs of $Z$ are called isolated, and ones of $U$ are called coupled. According to this, the symbols of $X^Z$ are called individual (special-purpose), and ones of $X^U$ are called general (general-purpose). 

In order to define a mechanism for coupled arcs we introduce the parent-arcs as a Cartesian product of child-arcs for state transition diagrams of subsystems of neighbour levels of hierarchy. The isolated arcs $Z$ represents the state transitions initiated by individual input symbols $X^Z$; this kind of symbols do not influence on the state transitions of other subsystems. The coupled arcs $U$ represents the state transitions initiated by general input symbols $X^U$; this kind of symbols initiate th state transition on the parent-arc, which means, as a consequence, the state transitions on the corresponding child-arcs. And conversely, state transitions on all or several child-arcs can initiate the state transition on the parent-arc of subsystem of higher level of hierarchy. 

A model of scenario controlling the development of control object is a 5-tuple
$$\left\{\Omega,I,M,C,V\right\}$$
$\Omega$ - a system of state transition diagrams; they represent the programs of state changes for each subsystem, \\
$I$ - the hierarchical structure, \\
$M:I\rightarrow \Omega$ - a functional that assignes a hierarchical number to each diagram of $\Omega$, \\
$C$ - time diagram for symbols $X$; it determines the sequential-parallel process of input symbols entering, \\
$V$ - a scheme of after-effect of state transitions.

To give time diagram $C$ of input control symbols entering, one can use various ways, including the estimation rules of each current state of system. 

The trajectory of attainable states represents general and local goals solved by scenario on arbitrary time interval. The study (investigation) basic properties of scenario is reduced to the analysis of trajectory of attainable states and its comparison with the expected or predicted effect. Some of the examples are:
\begin{enumerate}
	\item Completeness of scenario; this means the transition of all subsystems to  the final states of the corresponding state transition diagrams
	\item Redundancy of scenario; this means that the input symbols (signals) of different types, individual and general, enters at the input of a subsystem
	\item Omitted possibilities of scenario; this is exhibited by transition frequency on the arcs representing the backstep of the attained state; complexness of scenario in problem solving is estimated by transition frequency on the coupled arcs.
\end{enumerate}

\section{Conclusion}
\label{S4} \vspace{-4pt}

The presented models and analysis methods, and computer tools based on them, are the basis for development of applied systems for analysis, modeling and prediction of development processes of dynamical systems with the use of models of controllable development of hierarchical systems. The technique presented is also used as a technology for construction of information systems for simulation and analysis of development scenarios and strategies of complex objects, and has been applied in several information systems and decision support systems. 

We presented both general and special theory. The former concerns formalization of basic concepts and techniques for schematic representation and modeling of discrete hierarchical dynamic process; the latter one specialises the formalism to modeling the coordinating scenario-type control schemes. The method allows one to model inertial system dynamics that determines the current state consequences, and to demonstrate future state dynamics of system in arbitrary scenario-type control loop.

The information can be simultaneously aggregated in a few ways: by hierarchical structure of processes and states embedding, by parallel representation of dynamical characteristics of several processes within the framework of one state, and by dividing the observation time interval with regard to the events associated with the changes in system dynamics and tendencies. The proposed model is universal and at the same time it is problem-oriented in relation to the rationality and consistency of control actions; it can be equally used for diverse kinds of systems such as technical systems, organisational systems, systems of strategic planning and long-term forecasting systems, and decision support systems.

The main emphasis of the analysis methods of hierarchical model of state dynamics in control loop is on the efficiency of model of control scenario.

We suppose that theoretical model presented can serve as a basis for designing automated information systems for expert analysis and forecasting of complex large-scale system development.

\end{document}